\documentclass[fleqn,twoside]{article}
\usepackage{espcrc2}
\usepackage{psfig}

\newcommand{\AmS}{{\protect\the\textfont2
  A\kern-.1667em\lower.5ex\hbox{M}\kern-.125emS}}

\newcommand{\beeq}{\begin{equation}}
\newcommand{\eneq}{\end{equation}}
\newcommand{\beeqa}{\begin{eqnarray}}
\newcommand{\eneqa}{\end{eqnarray}}
\newcommand{\bma}{\begin{displaymath}}
\newcommand{\ema}{\end{displaymath}}

\title{
{\vspace{-1.2em} \parbox{\hsize}{\hbox to \hsize 
{\hss  \normalsize TRINLAT-02/01}}} \\
Supersymmetry on the lattice }

\author{Alessandra Feo\address[MCSD]{School of Mathematics, 
        Trinity College, Dublin 2, Ireland}
        \thanks{Plenary Talk at Lattice 2002, MIT, USA}
         } 

\begin{document}

\begin{abstract}
Lattice results in supersymmetry are summarized. Past, present and future perspectives are discussed.

\vspace{1pc}
\end{abstract}

\maketitle

\section{INTRODUCTION}
Supersymmetry or fermion-boson symmetry is one of the most fascinating topics in field theory. 
Even if has not yet been observed in Nature, thousands of papers have been written
on this subject. Its validity in particle physics follows from the common belief in unification
through the feasibility of incorporating quantum gravity.
From a theoretical point of view, non-perturbative studies of supersymmetric (SUSY) gauge theories 
turn out to have remarkably rich properties which are of great physical interest, as have been 
shown in \cite{seiberg}.
For this reason, much effort has been dedicated to formulating a lattice version of SUSY theories
(for a recent review of SUSY Yang-Mills (SYM) theories, using Wilson fermions, see \cite{montvay}).
The lattice formulation has been succesful to extract non-perturbative dynamics 
in field theory, specially in QCD, and may be able to provide additional information 
and confirm the existing analytical calculations.
Whether superymmetry is or not an exact symmetry is a question that must be settle by going beyond 
perturbation theory. 

\subsection{SYM theory in the continuum}
In the continuum, the action for the $N=1$ SYM theory with an $SU(N_c)$ gauge group is
\beeq
{\cal L} = -\frac{1}{4} F_{\mu \nu}^a(x) \, F_{\mu \nu}^a(x) + \frac{1}{2} \, \bar \lambda^a(x) \gamma_\mu 
{\cal D}_\mu \lambda(x)^a \, ,
\label{a1}
\eneq
where $\lambda^a$ is a 4-component Majorana spinor and satisfies the Majorana condition 
$\bar \lambda^a = {{\lambda}^{a}}^T C$. The gluon fields are represented by
$A_\mu = -i g A_\mu^a T^a $ and $ F_{\mu \nu} = -i g F_{\mu \nu}^{a} T^a$.
${\cal D}_\mu \lambda^a = \partial_\mu \lambda^a + g f_{abc} A_\mu^b \lambda^c$
is the covariant derivative in the adjoint representation.
The continuum SUSY transformations read
\beeqa
&& \hspace{-0.7 cm} \delta A_\mu(x) = -2 g \bar\lambda(x) \gamma_\mu \varepsilon \, ,  \\
&& \hspace{-0.7 cm} \delta \lambda(x) = -\frac{i}{g}\sigma_{\rho\tau} F_{\rho\tau}(x) \varepsilon \, , \, \, \, \,
\delta \bar\lambda(x) = \frac{i}{g} \bar \varepsilon \sigma_{\rho\tau}F_{\rho\tau}(x) \, , \nonumber
\label{a5}
\eneqa
where $\sigma_{\rho\tau} = \frac{i}{2} [\gamma_\rho,\gamma_\tau] $, $\lambda = \lambda^a T^a$ 
and $\varepsilon $ is a global Grassmann parameter with Majorana properties.
These transformations relate fermions and bosons, leave the action invariant
and commute with the gauge transformations so that the resulting Noether current $S_\mu(x)$ is 
gauge invariant. For $N=1$ SYM theory the supercurrent is given by
\beeq
S_\mu(x) = - F_{\rho \tau}^a(x) \sigma_{\rho \tau} \gamma_\mu \lambda^a(x) \, .
\label{a12}
\eneq
Classically, the Noether current is conserved, $\partial_\mu S_\mu(x) = 0$, provided the fields satisfy
the equations of motion. Furthermore one finds that it fullfills a spin $3/2$ constraint:
$\gamma_\mu S_\mu(x) = 0$. 

\subsection{Superfields}
Supersymmetry in Eq.~(\ref{a1}) can be better understood in the language of superfields \cite{bagger,weinberg}.
In the Wess-Zumino gauge, the action for the $N=1$ SYM theory is 
\beeq
S_{SYM} = \mbox{Re} \, \bigg\{ \int d^4 x \, d^2 \theta \, \mbox{Tr} [W^\alpha W_\alpha] \bigg\} \, .
\eneq
$W(x,\theta,\bar \theta)_\alpha = -\frac{1}{8} (\bar D \, \bar D) e^{-2 V} D_\alpha e^{2 V}$
is the spinorial field strength superfield which depends on the four 
coordinates $x$ and the anticommuting Weyl spinor variables $\theta_\alpha$, $\bar \theta_{\dot{\alpha}}$ 
$(\alpha, \dot{\alpha} =1,2)$. 
$\bar D_{\dot{\alpha}}$ and $D_\alpha$ are the superspace derivatives while $V=V(x,\theta,\bar \theta)$ is the 
vector superfield which takes the form \cite{bagger}
\beeqa
V  &=& \theta \sigma^\mu \bar \theta A_\mu(x) + 
i (\theta \theta) \bar \theta \bar \psi_w(x) \nonumber \\ 
&& - i (\bar \theta \bar \theta) \theta \psi_w(x) + \frac{1}{2} (\theta \theta) (\bar \theta \bar \theta) D(x) \, .
\eneqa
The particle content can be identified as one bosonic vector field $A_\mu = -i g A_\mu^a T^a$, 
its SUSY partner, a complex Weyl spinor field $\psi_w = T^a \psi_w^a$, and an auxiliary scalar field $D$. 
The action can be expressed in terms of these components as 
\beeqa
&& \hspace{-0.7 cm} S_{SYM} = \int d^4x \Bigg\{ -\frac{i}{4} F_{\mu \nu}^a F^{\mu \nu a}
+ \frac{1}{2} i \psi_w^a \sigma^\mu ({\cal D}_\mu \bar \psi_w)^a \nonumber \\
&&  - \frac{1}{2} i ({\cal D}_\mu \bar \psi_w)^a \bar \sigma^\mu \psi_w^a + \frac{1}{2} D^2 \Bigg\} \, .
\label{a7}
\eneqa
In the on-shell case, the auxiliary scalar field $D$ is eliminated by the equation of motion $D=0$.
Introducing one massless Majorana fermion, the gluino, in the adjoint representation 
$\lambda = ((\psi_w)_\alpha,(\bar \psi_w)^{\dot{\alpha}})$, and going to the Euclidean space, 
we recover Eq.~(\ref{a1}).

\subsection{SUSY Ward Identities (WIs)}
The existence of a renormalized SUSY Ward-type identity
\beeq
\partial_\mu S_\mu^R(x) = 2 m_R \chi_R(x) \, ,
\label{a10}
\eneq
is generally assumed. It is obtained by performing a variation of the functional integral with respect to a 
local, smooth transformation, $\varepsilon = \varepsilon(x)$, and then putting the sources to zero.  
$S_\mu^R$ is the renormalized supercurrent and $\chi_R = Z_\chi \chi$, with
$\chi \equiv \frac{1}{2} F_{\mu \nu}^a \sigma_{\mu \nu} \lambda^a $.
$m_R$ is the renormalized gluino mass. For $m_R = 0$ we have supersymmetry while a  
non-vanishing value of $m_R$ breaks supersymmetry softly. 

\subsection{Non-perturbative effects of the SYM dynamics}
Because of asymptotic freedom and the corresponding infrared instability, it is reasonable to guess
that, as in conventional QCD, there are many non-perturbative phenomena that occur in this theory.
For the $N=1$ SYM theory we may expect confinement and spontaneous chiral symmetry breaking.
The discrete chiral symmetry is expected to be broken by a non-zero gluino condensate
while the confinement is realized by colorless bound states described by the effective action
belonging to chiral supermultiplets.

It is generally assumed that supersymmetry is not anomalous (Eq.~(\ref{a10}) holds) and only 
the mass term is responsible for a soft breaking. 
However, in \cite{shamir} the question of whether non-perturbative effects may cause a supersymmetry anomaly 
has been raised. 
Only a study of the continuum limit of the lattice SUSY WIs can shed light on this question.

\subsection{Chiral symmetry breaking}
Introducing a non-zero gluino mass term in Eq.~(\ref{a1}), 
${\cal L}_{mass} = m_{\tilde g}\bar \lambda^a \lambda^a$, 
breaks supersymmetry softly (which implies that the non-renormalization theorem and cancellation
of divergencies are preserved \cite{girardello}).
In the massless case, the global chiral symmetry is $U(1)_\lambda$: 
\beeq
\lambda \to e^{-i \varphi \gamma_5} \lambda \, , \qquad \qquad 
\bar \lambda \to \bar \lambda e^{-i \varphi \gamma_5} \, ,
\label{a6}
\eneq
which is anomalous. This implies that the divergence of the axial current,
$J^5_\mu = \bar \lambda \gamma_5 \lambda$, is
\beeq
\partial_\mu J^5_\mu = \frac{N_c g^2}{32 \pi^2} \varepsilon^{\mu \nu \rho \sigma} F_{\mu \nu}^a 
F_{\rho \sigma}^a \, .
\eneq
The anomaly leaves a $Z_{2 N_c}$ subgroup of $U(1)_\lambda$ unbroken.
Eq.~(\ref{a6}) is equivalent to 
$m_{\tilde g} \to m_{\tilde g} e^{-2 i\varphi \gamma_5}$ and 
$\Theta_{SYM} \to \Theta_{SYM} - 2 N_c \varphi$.

In the SUSY case, $m_{\tilde g}=0$, the $U(1)_\lambda$ symmetry is unbroken if
$\varphi \equiv \frac{k \pi}{N_c}$ for $(k=0,1,\cdots,2 N_c -1)$.
$Z_{2N_c}$ is expected to be spontaneously broken to $Z_2$ by a value of 
$\big< \bar \lambda \lambda \big> \not = 0$ \cite{witten}. The consequence of this 
spontaneous chiral symmetry breaking is the existence of a first order phase
transition at $m_{\tilde g} = 0$. That means the existence of $N_c$ degenerate ground
states with different orientations of the gluino condensate $(k=0,\cdots,N_c -1)$,
\beeq
\big<\bar \lambda \lambda \big> = c \Lambda^3 e^{\frac{2 \pi i k}{N_c}} \, ,
\label{a2}
\eneq
where $\Lambda$ is the dynamical scale of the theory which can be calculated on the lattice, for example, 
while $c$ is a numerical constant which depends on the renormalization scheme 
used to compute $\Lambda$.
Eq.~(\ref{a2}) shows the dependence on the gauge group.  
For $SU(2)$ two degenerate ground states with opposite sign of the
gluino condensate, $\big< \bar \lambda \lambda \big> < 0$ and 
$\big< \bar \lambda \lambda \big> > 0$ appear, while for $SU(3)$ there 
are three degenerate vacua at $k =k_c$ (in \cite{kovner} a fourth state 
with $\big<\lambda \lambda \big>=0$ is claimed). 
A first numerical study of SYM theory with a gauge group $SU(3)$ is in \cite{feo3}.

\subsection{Magnitude of the gluino condensate}
The calculation of the gluino condensate for the $N=1$ SYM theory is a puzzle.
Two approachs in the literature have been used which give different results for $c$.
One is based on weak-coupling instanton (WCI) calculations \cite{affleck}
and gives $c=1$. In the second, calculations based on strong-coupling instanton (SCI) give 
$c = 2/((N_c - 1)! (3 N_c -1))^{1/N_c}$ \cite{veneziano}. For a review on SCI see
\cite{amati}.
Several suggestions have been put forward to resolve the puzzle, 
see for example \cite{ritz}. Ref.~\cite{hollowood} cast serious doubts on the SCI 
calculation by showing that the cluster decomposition is not valid. 
It is also shown that the addition of the so-called Kovner-Shifman (KS) chiral symmetric vacuum state 
\cite{kovner,vainshtein} can not straightforwardly resolve the disagreement between
the SCI and WCI results \cite{hollowood}. The KS vacuum can indeed potentially resolve
the mismatch at the 1-instanton sector $(k=1)$ but it fails to do so for the 
topological sectors with $k>1$ \cite{hollowood}.
Using an instanton liquid picture gives qualitatively similar results and evidence 
for the gluino condensate \cite{schafer}.

\subsection{Light hadron spectrum}
For the $N=1$ SYM theory a low energy effective action has been proposed by
Veneziano and Yankielowicz (VY) \cite{veneziano2}. The action contains all degrees of 
freedom, gauge invariant and colorless, composite fields:
$F^a_{\mu \nu} F^a_{\mu \nu}$,
$F^a_{\mu \nu} \tilde{F}^a_{\mu \nu}$,
$\bar \lambda^a \lambda^a$, in analogy to QCD, while 
$\chi = \sigma_{\mu \nu } F^a_{\mu \nu} \lambda^a $ is a new type of composite operator formed 
by a gluino $\lambda$ and a gauge field $F$, both in the adjoint representation.  
These fields can be combined to form the chiral supermultiplet, containing 
the expression for the anomalies as component fields \cite{ferrara}
\beeq
S(x,\theta) = \phi(x) + \sqrt{2} \theta \chi(x) + \theta \theta F(x) \, .
\eneq
$\phi = \frac{\beta(g)}{2 g}(\psi_w)^\alpha (\psi_w)_\alpha$, is proportional to the 
gluino bilinear (here $\psi_w$ denotes a 2-component Weyl spinor), while 
the other components, which we do not report here (see \cite{bagger}),
contain combinations of gluino-gluino field and gluino-gluon fields. 
The VY action has the form \cite{bagger}
\bma
{\cal L}_{eff}= \frac{1}{\alpha} (S^\dagger S)^{1/3} \vert_D + \gamma [(S \mbox{log} \frac{S}{\mu^3}
- S) \vert_F + h.c. ] \, .
\ema
Expanding the effective VY action around its minimum, it is found that the low energy 
spectrum forms a supermultiplet, consisting of a scalar meson $\bar \lambda^a \lambda^a$
(the $a-f_0$), a pseudoscalar meson $\bar \lambda^a \gamma_5 \lambda^a$ (the $a-\eta \prime$), 
(the $a$ denoting the adjoint representation),
and a spin-$1/2$ gluino-glueball particle, the $\chi$. 
Glueballs are absent in this formulation.

In the SUSY point these masses are degenerate. The introduction of a 
$m_{\tilde{g}} \ne 0$ breaks supersymmetry softly and leads to
a splitting of the multiplet. How the spectrum is influenced by the soft SUSY breaking 
has been studied in \cite{evans}:
\beeqa
M_{a - \eta \prime} &= & N_c \alpha \Lambda + \frac{40 \pi^2 \vert m_{\tilde{g}} \vert}{3 N_c} + \cdots \, , \nonumber \\
M_{a - \chi} &= & N_c \alpha \Lambda + \frac{48 \pi^2 \vert m_{\tilde{g}} \vert}{3 N_c} + \cdots \, , \nonumber \\
M_{a - f_0} &= & N_c \alpha \Lambda + \frac{56 \pi^2 \vert m_{\tilde{g}} \vert}{3 N_c} + \cdots \, .
\label{a3}
\eneqa
Unfortunately, the range of applicability of the linear mass formulae is not known 
because of the unknown magnitude of the constants and of the higher
order terms in Eq.~(\ref{a3}).

Introducing an extra term in the effective VY action, Farrar {\em et al.} (FGS) 
solved the question of including glueballs in the low energy spectrum \cite{farrar}.
For unbroken supersymmetry we expect to see two chiral multiplets (not one), at the bottom
of the SYM spectrum.
The lighter one contains a $0^+ \mbox{glueball} \approx F_{\mu \nu} F_{\mu \nu} $,
a $0^- \mbox{glueball} \approx \varepsilon^{\mu \nu \rho \sigma} F_{\mu \nu} F_{\rho \sigma} $ 
and a gluino-glueball ground state, while the heavier supermultiplet is the VY one.
Moreover, in the FGS picture, a non-zero mixing between the $a-f_0$ and $0^+$ glueball is possible. 
Of course, other generalizations beside the FGS picture are conceivable.

\section{LATTICE FORMULATION OF $N=1$ SYM }
The question of whether it is possible to formulate successfully SUSY theories on the lattice 
has been addressed in the past by several authors \cite{curci,dondi,golterman} 
(the reader is refered to \cite{golterman} for a sucessful contruction 
of a lattice $2d$ Wess-Zumino model, and some discussion concerning Ref.~\cite{dondi}).
It can be seen that a lattice regularized version of a gauge theory is not SUSY 
since the Poincar\'e invariance (a sector of the superalgebra) is lost, thus
$\{ Q_\alpha,\bar Q_\beta \} = 2 \sigma^\mu_{\alpha \beta} P_\mu $. 
Poincar\'e invariance is achieved automatically without fine tuning in the continuum limit
because operators that violate Poincar\'e invariance are all irrelevant.
Moreover, if the SUSY theory contains scalar fields one can have scalar mass terms that break supersymmetry.
Since these operators are relevant, fine tuning is necessary in order to cancel their contributions.

Another problem is the question of how to balance bosonic and fermionic modes, the numbers
of which are constrained by the supersymmetry: the naive lattice fermion 
formulation results in the doubling problem \cite{nielsen},
and produces a wrong number of fermions. The problem can be treated as in QCD
by using different fermion formulations. Let us briefly summarize 
those which have applications in SUSY theories.

\subsection{Wilson fermions}
In the Wilson formulation for the $N=1$ SYM theory it is proposed to give up manifest
supersymmetry on the lattice and restore it in the continuum limit \cite{curci}.
Supersymmetry is broken by the lattice, by the Wilson term and is softly broken by the presence of
the gluino mass. Supersymmetry is recovered in the continuum limit by tuning the bare 
parameters $g$ and the gluino mass $m_{\tilde{g}}$ to the SUSY point. The chiral and 
SUSY limit can be recovered simultaneously at $m_{\tilde{g}}=0$.

In the Wilson formulation, the Curci and Veneziano (CV) effective action, suitable for Monte Carlo 
simulations, reads
\bma
S_{CV} = \beta \sum_{pl} \bigg( 1 - \frac{1}{N_c} \mbox{Re} \mbox{Tr} U_{\mu \nu} \bigg) - \frac{1}{2} 
\mbox{log} \, \mbox{det} Q[U] \, . 
\ema
For the gauge group $SU(N_c)$, the bare coupling is given by $\beta \equiv 2 N_c/g^2$. 
The fermion matrix for the gluino $Q$ is defined by
\beeqa
\label{a15}
&& \hspace{-0.6 cm} Q_{yb,xa}[U] \equiv 
  \delta_{xy} \delta_{ab} - k \sum_{\mu=1}^4 \big[ \\
&& \hspace{-0.3 cm} \delta_{y, x + \hat \mu} (1 + \gamma_\mu) V_{ba,x \mu} + 
   \delta_{y + \hat \mu, x} (1 - \gamma_\mu) V^T_{ba,x \mu} \big] \, . \nonumber 
\eneqa
$k$ is the hopping parameter defined as $k = 1/(2(4 + m_0 a))$, where $m_0$ is the bare mass,
and the matrix for the gauge field link in the adjoint representation is
\beeq
V_{ab,x \mu} \equiv  V_{ab,x \mu}[U] \equiv 
\frac{1}{2} \mbox{Tr}( U^\dagger_{x \mu} T_a U_{x \mu} T_b ) \, .  
\label{a9}
\eneq
The fermion matrix for the gluino in Eq.~(\ref{a15}) is not hermitian but it satisfies
$Q^\dagger = \gamma_5 Q \gamma_5$. That allows for the definition of the hermitian
fermion matrix $\tilde Q \equiv \gamma_5 Q$. 
The path integral over the Majorana fermions gives the Pfaffian,
\bma
\int [d \lambda] e^{-\frac{1}{2} \bar \lambda Q \lambda} =
\int [d \lambda] e^{-\frac{1}{2} \lambda^T C Q \lambda} = Pf(M) \, ,
\ema
where $M \equiv C Q$ is an antisymmetric matrix.

It is easy to see that $Pf(M) = \pm\sqrt{det \, Q}$. In the effective CV action the absolute
value of the Pfaffian is taken into account (this may cause the sign problem). 
The omitted sign can be included by reweighting the expectation values according to the formula,
\beeq
\big< {\cal O} \big> = \frac{\big< {\cal O} \, \mbox{sign}Pf(M) \big>_{CV}}{\big< \mbox{sign}Pf(M) \big>_{CV}} \, .
\eneq
The spectral flow is a method which checks the value of the sign of the Pfaffian. The experience
of the DESY-M\"unster collaboration shows that bellow the critical line 
$k_c(\beta)$, corresponding to zero gluino mass $(m_{\tilde{g}} =0)$, negative Pfaffians 
practically never appear \cite{feo,feo2,feo4}.

The factor $1/2$ in front of $\mbox{log}\, \mbox{det} \, Q[U]$ shows that Majorana fermions
imply a flavor number $N_f = 1/2$. This can be achieved by the hybrid molecular dynamics
(HMD) algorithm \cite{gottlieb} which is applicable to any number of flavors.
The HMD has been checked for the CV action in $N=1$ $SU(2)$ SYM at small lattices 
$4^3 \times 8$ \cite{donini}.
Another method for simulation with non-even numbers of flavors is 
based on the multi-bosonic algorithm proposed by L\"uscher \cite{luscher}. 
A two-step variant using a noisy correction step \cite{kennedy},
has been developed by Montvay in \cite{montvay2,montvay3} called the two-step 
multibosonic (TSMB) algorithm.
In the two-step variant, to represent the fermion determinant one uses a first polynomial 
${\cal P}_{n_1}^{(1)}(x)$ for a crude approximation realizing a fine correction by another polynomial
${\cal P}_{n_2}^{(2)}(x)$ that satisfies, 
${\displaystyle \lim_{n_2 \to \infty}} {\cal P}_{n_1}^{(1)}(x) {\cal P}_{n_2}^{(2)}(x) = x^{-N_f/2}$,
for $x \in \, [\varepsilon,\lambda]$.
The fermion determinant is approximated as \cite{montvay2}
\bma
\mbox{det}(Q^\dagger Q)^{N_f} \simeq \frac{1}{\mbox{det} \, P_{n_1}^{(1)}(Q^\dagger Q) 
\mbox{det} \, P_{n_2}^{(2)}(Q^\dagger Q)} \, .
\ema
Unquenched results using TSMB have been reported by the DESY-M\"unster-Roma collaboration
(the first large scale numerical simulation of $N=1$ SYM theory). For $SU(2)$ see
\cite{kirchner,feo,feo2}, while for $SU(3)$ see \cite{feo3}.
Interesting quenched results (which was pioneering work) are in \cite{koutsoumbas,donini2,vladikas}.

\begin{figure}[tb]
\null\vskip2mm
\hspace{-0.3 cm}\centerline{
          \psfig{figure=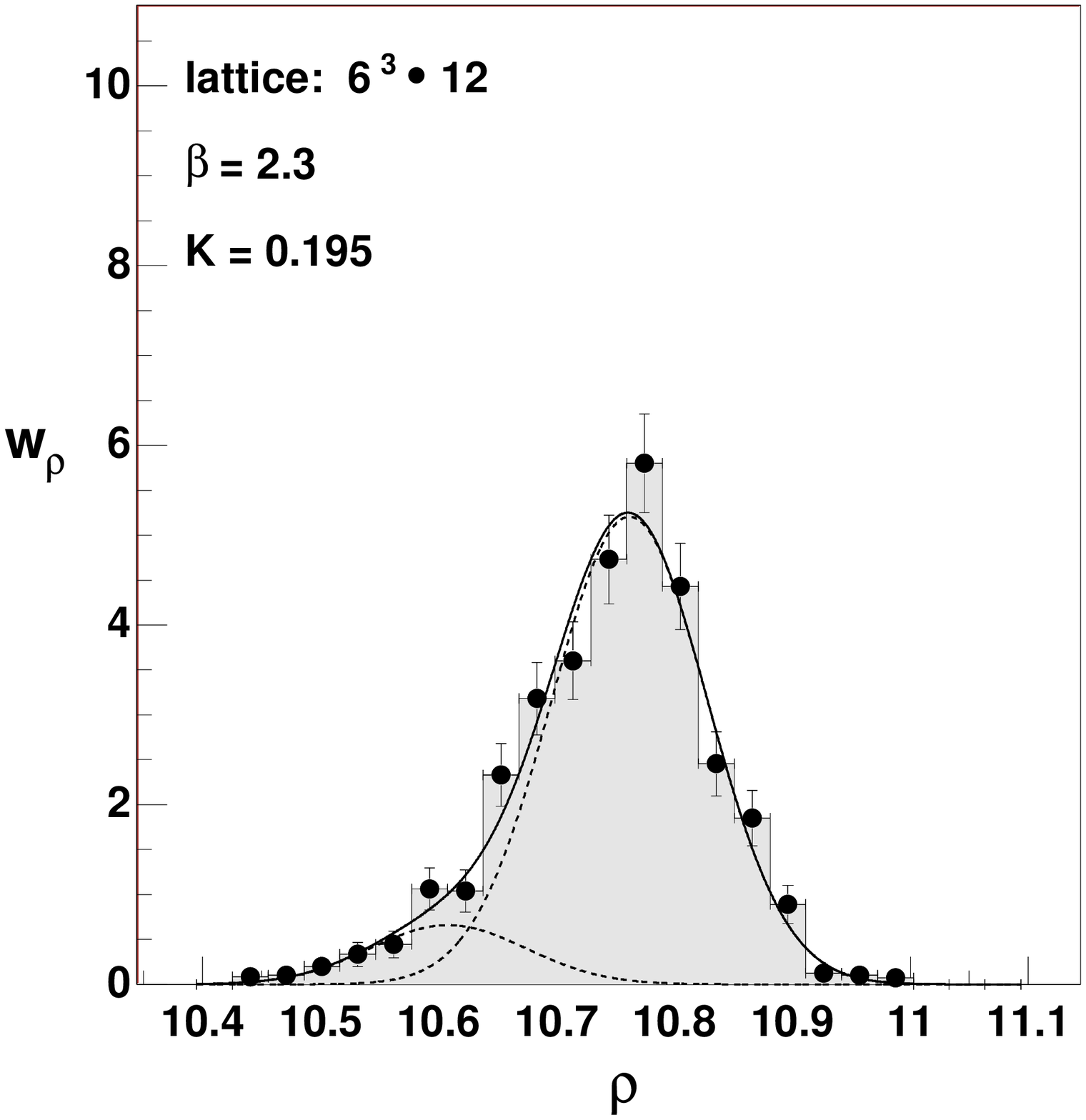,height=3.8cm}
          \psfig{figure=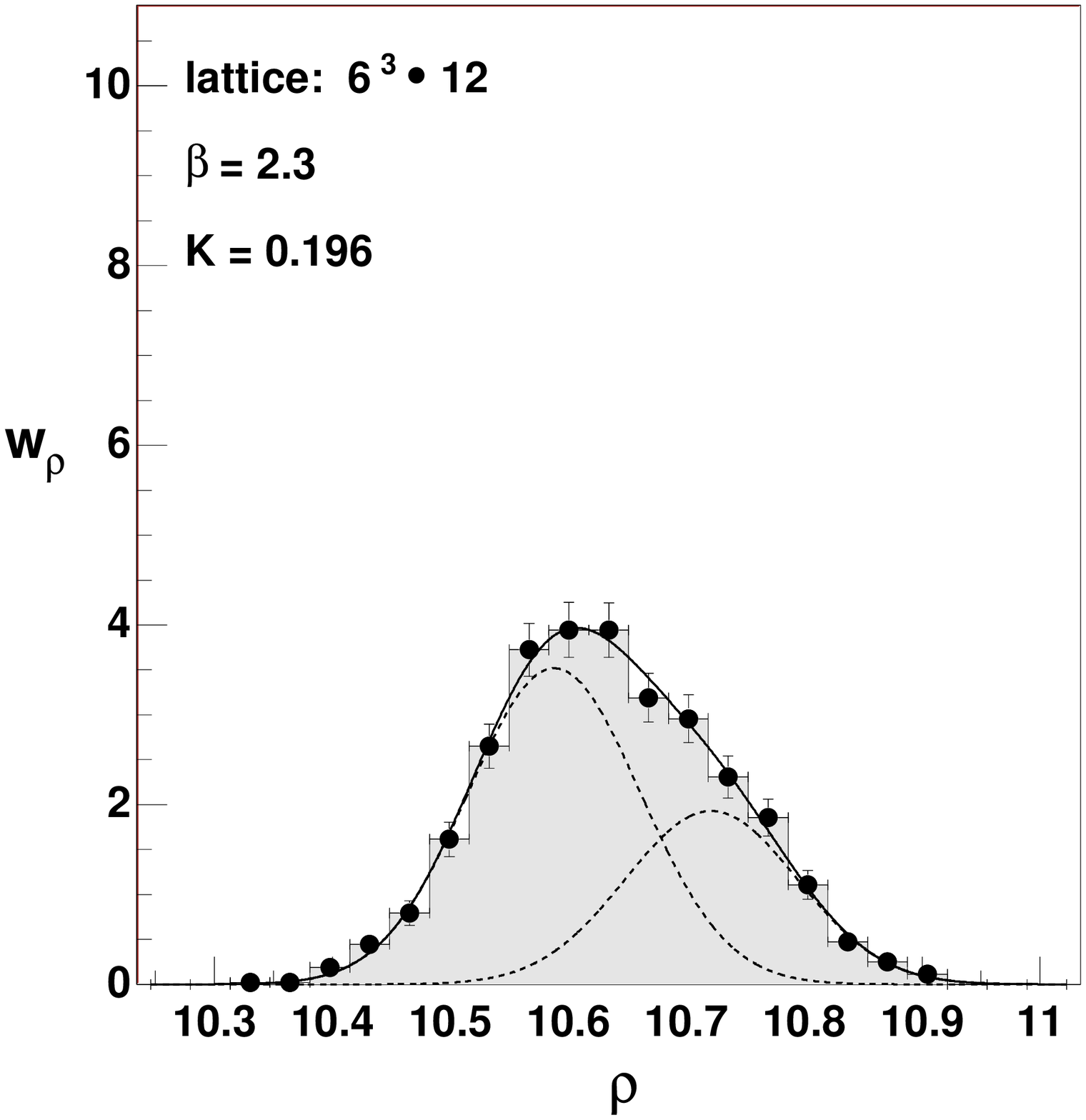,height=3.8cm}}
\vskip-8mm
\caption{Distribution of the gluino condensate for $k = 0.195$ and $k=0.196$.
$k_c = (0.1955 \pm 0.0005)$.}
\label{fig.1}
\end{figure}

In order to check the expected pattern of spontaneous chiral symmetry breaking, 
let us first write the expression for the renormalized gluino condensate, obtained
by additive and multiplicative renormalizations: 
$\big< \bar \lambda \lambda \big>_{R(\mu)} = Z(a \mu) [\big< \bar \lambda \lambda \big> - 
b_0(a \mu)]$. 
In a numerical simulation, a first order phase transition (or cross-over) should show up 
(on small or modere lattices) as a jump in the expectation value of the gluino condensate
at $k = k_c$. By tuning the hopping parameter $k$ to $k_c$, for a fixed $\beta$ value, 
one expects to see a two peak structure in the distribution of the gluino condensate.
By increasing the volume the tunneling between the two ground states becomes less and less 
probable and at some point practically impossible. Outside the phase transition region,
the observed distribution can be fitted by a single Gaussian but in the transition region 
a good fit can only be obtained with two Gaussians. 
The DESY-M\"unster collaboration performed the first lattice investigation of the gluino condensate. 
Results for $SU(2)$ are reported in Fig.~\ref{fig.1} and show that a first order phase transition 
occurs at $k_c = (0.1955 \pm 0.0005) $ for $\beta = 2.3 $. The lattice size used is $6^3 \times 12$ 
\cite{kirchner}. For $SU(3)$, preliminary results are in \cite{feo3}. 
Here the pattern is more complicated, $Z_6 \to Z_2$. The probability distribution for the gluino
condensate was measured for $k=0.1950$ and $\beta =5.6$ on a rather small lattice $4^3 \times 8$, with 
encouraging results.
Of course, a more accurate study by going to larger lattices will clarify the nature of the transition.

SUSY restoration can be also verified by a direct inspection of the low energy mass spectrum
\cite{feo,feo8,montvay}: this is expected to reproduce the SUSY multiplets 
predicted in \cite{veneziano2,farrar}. An accurate analysis of the spectrum is, however,
a non-trivial task from the computational point of view and an independent method for 
checking SUSY restoration would be welcome. Moreover, the mixing between the states has been measured.
Numerical simulations show that this mixing is practically zero \cite{feo,kirchner2}. 
At this conference, new results for the low lying spectrum of $N=1$ SYM, for 
$12^3 \times 24$ and $k=0.1955$ and for $16^3 \times 32$ and $k = 0.194$, are reported by Peetz \cite{peetz}. 

A study with a large number of colors $N_c$ and strong coupling, by considering the hopping 
parameter expansion as a sum over lattice paths (random walks), is \cite{arroyo}. 
The surprising result is that the quenched approximation is exact at this order (and it is also
consistent with the quenched results of Donini et al. in \cite{donini2}).
Formulae for the propagators and masses of 2- and 3-gluino states are also presented.
The 2-gluino masses do coincide with the results for the meson spectrum in ordinary lattice
QCD at strong coupling \cite{arroyo}.
A preliminary study of a 3-gluino states is reported also in \cite{kirchner2,kirchner3}.
This particle does not appear in \cite{veneziano2,farrar} even if it contains the same
quantum numbers. A numerical analysis of this issue would clarify whether it
contributes or not in the mass spectrum.
At this conference, a study of the strong coupling expansion for $N=1$ SYM theory 
using the Hamiltonian formalism is also presented \cite{berruto}.
It is shown that the theory is effectively described by an $SO(4)$ antiferromagnet.

\subsection{Domain wall fermions}
A very nice innovation is the domain wall fermions (DWF) approach. A new lattice fermion regulator,
which improves the lattice formulation for fermions because the zero gluino mass is achieved without
fine tuning.
Application of DWF in SUSY theories has been explored in \cite{neuberger,kaplan}
and also suggested in \cite{nishimura}, with a different approach as \cite{kaplan}.
First Monte Carlo simulations for $N=1$ $SU(2)$ SYM with DWF, using the lines of 
Refs.~\cite{neuberger,kaplan} are in \cite{vranas,fleming}.
DWF were introduced in \cite{kaplan2} and were further developed in \cite{narayanan,shamir2}.
For a recent review on DWF for SUSY gauge theories see \cite{vranas2}. 

There are two unwelcome difficulties in using Wilson fermions as mentioned in the previous section. 
The first one is the need for fine tuning. The second one is related with the sign of the Pfaffian.
DWF are defined by extending the space-time to five dimensions. Also a non-zero five dimensional mass
or domain wall height $m_0$, which controls the number of flavors, is present. 
$L_s$ is the size of the fifth dimension and free boundary conditions for the 
fermions are implemented.
As a result the two chiral components of the Dirac fermion are separated with one chirality 
bound exponentially on one wall and the other on the opposite wall.
For any value of $a$ the two chiralities mix only by an amount that decreases exponentially 
as $L_s \to \infty$. For $L_s = \infty$, chiral symmetry is expected to be exact even at finite lattice spacing.
So, there is no need for fine tuning \cite{vranas2}.
DWF offer for the first time the oportunity to separate the continuum limit, $a \to 0$, 
from the chiral limit, $L_s \to \infty$.

DWF introduce two extra parameters: $L_s$ and $m_0$. 
These two parameters together with the four dimensional mass $m_f$ control the effective
fermion mass $m_{eff}$. In the free theory one finds \cite{vranas3} 
\beeq
m_{eff} = m_0 (2-m_0) [ m_f + (1 - m_0)^{L_s}] \, , 
\label{a16}
\eneq
with $0 < m_0 < 2$. The value of $m_0 = 1$ is optimal because finite $L_s$ effects do not contribute 
to $m_{eff}$. In the interactive theory, one would not expect such optimal value, due to the fact that $m_0$ will 
fluctuate. Then the goal would be to have $L_s$ large enough to have the second term of Eq.~(\ref{a16})
small, in order to simulate at reasonably small masses and extrapolate to the chiral limit,
$m_f \to 0$ and $L_s \to \infty$ \cite{vranas}.

The effective action, which is not reported here (more details in \cite{vranas}) contains $L_s$
heavy species. When $L_s$ is increased they may introduce bulk effects which must be substracted.
This can be done by introducing Pauli-Villars type fields \cite{narayanan}.
The fermionic path integral gives the Pfaffian 
\beeq
\int [d \lambda] e^{-S_F} = Pf(M_F) = \sqrt{det(D_F)} 
\eneq
which is positive for $m_f > 0$. This is a second advantage of the DWF approach.

\begin{figure}[tb]
\null\vskip2mm
\centerline{\psfig{figure=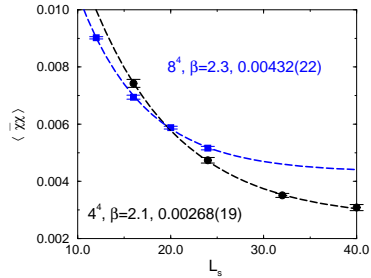,height=4cm}}
\vskip-8mm
\caption{Dynamical gluino condensate at $m_f=0$ vs $L_s $ on two different lattices.}
\label{fig.2}
\end{figure}

Numerical simulations with DWF for $N=1$ $SU(2)$ SYM are reported in \cite{vranas,fleming}, using 
the (HMDR) of \cite{gottlieb}.
They concern only the study of the gluino condensate, for rather small volumes, $4^4$ and $8^4$. 
The results for $m_f =0$ are shown in Fig.~\ref{fig.2}. 
As can be seen, $\big<\bar \lambda \lambda \big> (L_s \to \infty)$
has a non-zero VEV which remains different from zero, even for rather small volumes.
These results (and also those in \cite{vranas}) support the non-zero value for the gluino 
condensate.

Also the DWF has difficulties in its implementation. Beside the two extra parameters,
it may happen that the two chiralities do not decouple, even for $L_s \to \infty$. 
In this case the chiral symmetry can not be restored.
This may need large values of $L_s$ and for this reason much expensive cost in the simulations. 
From the computational point of view, SYM is harder to simulate than QCD, using DWF, 
while with Wilson fermions, SYM is easier to simulate than QCD.

\subsection{SUSY WIs on the lattice}
Another independent way to study the SUSY limit is by means of the SUSY WIs.
On the lattice they contains explicit SUSY breaking terms and 
the SUSY limit is defined to be the point in parameter space where these breaking terms
vanish and the SUSY WIs take their continuum form. These issues have been investigated numerically by 
the DESY-M\"unster-Roma collaboration, using Wilson fermions \cite{feo2,feo4}.  
Previoulsy in \cite{vladikas}. 

In the continuum, the SUSY WIs are given by Eq.~(\ref{a10}) where 
$S_\mu^R(x) \equiv Z_S S_\mu(x) + Z_T T_\mu(x)$ 
and $Z_S$ and $Z_T$ are multiplicative renormalization factors.
$T_\mu(x) = F_{\mu \nu}^a(x) \gamma_\mu \lambda^a(x)$ is a mixing current with dimension $7/2$
as $S_\mu(x)$ \cite{feo2}.
 
In the Wilson formulation, supersymmetry is not realized on the lattice. One might still define some 
lattice SUSY transformations (which reduce to the continuum ones (\ref{a5}) in the limit $a \to 0$). 
One choice is \cite{feo2,taniguchi}
\beeqa
\delta U_\mu(x) & = & -
a g U_\mu(x) \bar \varepsilon(x) \gamma_\mu \lambda(x) \nonumber \\
&& - a g \bar \varepsilon(x + a \hat\mu) \gamma_\mu \lambda(x + a \hat\mu) U_\mu(x) \, , \nonumber \\
\delta \lambda(x)& =&
- \frac{i}{g} \sigma_{\rho\tau} {\cal G}_{\rho \tau}(x) \varepsilon(x) \, . 
\eneqa
In the case of a gauge invariant operator insertion ${\cal O}(y)$, we find for the bare SUSY WIs 
\beeqa 
&& \hspace{-0.7 cm} \big< {\cal O}(y) \nabla_\mu S_\mu(x) \big>  -
2 m_0 \big< {\cal O}(y) \chi(x) \big> + 
\big< \frac{\delta{\cal O}(y)}{\delta \bar \varepsilon(x)}|_{\varepsilon = 0} \big>  \nonumber \\ 
&& = \big< {\cal O}(y) X_S(x) \big> \, .
\eneqa
Comparing with Eq.~(\ref{a10}), beside the presence of a non-zero bare mass term in the action which
breaks supersymmetry soflty, the rest of the SUSY breaking results in the presence of the 
$X_S$ term, that appear since the action is not fully SUSY.
In order to renormalize the SUSY WIs a possible operator mixing has to be taken into
account. In the case of gauge invariant operator insertion, 
$X_S$ mixes with the following operators of equal or lower dimension \cite{bochicchio},
$\nabla_\mu S_\mu $, $ \nabla_\mu T_\mu $ and $\chi $.
The SUSY WIs can be written as \cite{vladikas}
\beeqa
&& \hspace{-0.7 cm} \big< {\cal O}(y) \nabla_\mu S_\mu(x) \big> +
Z_T Z_S^{-1} \big< {\cal O}(y)\nabla_\mu T_\mu(x) \big>   \nonumber \\
&& = m_R Z_S^{-1} \big< {\cal O}(y) \chi(x) \big> \, .
\label{a13}
\eneqa
Here the gauge invariant operator ${\cal O}(y)$ at point $y$ is assumed to be sufficiently 
far away from $x$ in such a way that contact terms are avoided. This is the on-shell regime.
$m_R = m_0 - a^{-1} Z_\chi$ and $\nabla_\mu$ is the lattice derivative.
In numerical simulations, Eq.~(\ref{a13}) can be computed at fixed $\beta$ 
and $k$. Thus, by choosing two elements of the $4 \times 4$ matrices, a system of two equations
can be solved for $Z_T Z_S^{-1}$ and $m_R Z_S^{-1}$.
One must clearly ensure that these two equations are non-trivial and independent. To do this,
a proper definition of the supercurrent operator and the operator insertion ${\cal O}$ is necessary.
In Ref.~\cite{feo2}, two different definitions for the supercurrent operator are considered in Eq.~(\ref{a13}): 
the local definition (\ref{a12}), and more involved, the point-split definition, reported also  
in \cite{taniguchi}, which differs from the local one in order ${\cal O}(a)$.
  
\begin{figure}[tb]
\null\vskip2mm
\hspace{-0.1 cm} \centerline{\psfig{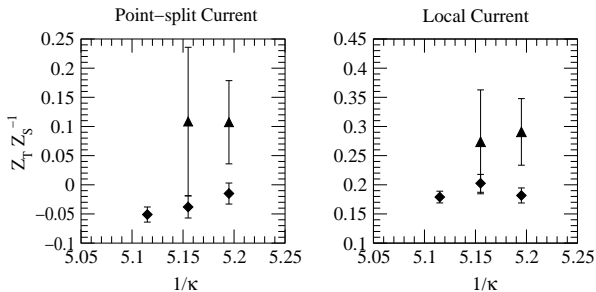}}
\vskip-8mm
\caption{$Z_T Z_S^{-1}$ as a function of $1/k$ with the operator insertion
$\chi(x)$ (filled diamonds) and $T_0(x)$ (filled triangles).}
\label{fig.3}
\end{figure}

\begin{figure}[tb]
\null\vskip2mm
\hspace{-0.1 cm} \centerline{\psfig{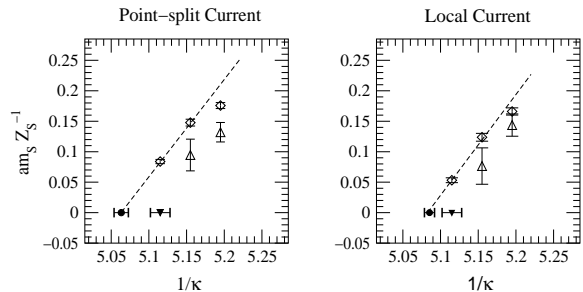}}
\vskip-8mm
\caption{$am_s Z_S^{-1}$ as a function of $1/k$ with the operator insertion
$\chi(x)$ (diamonds) and $T_0(x)$ (triangles).}
\label{fig.4}
\end{figure}

The dependence of $Z_T Z_S^{-1}$ and $m_R Z_S^{-1}$ on $k$, are shown in Fig.~\ref{fig.3}
and Fig.~\ref{fig.4}, respectively, with two different operator insertion, 
$\chi(x)$ and $T_0(x)$, for a lattice size $12^3 \times 24$. 
Fitting results are considered only for the insertion $\chi(x)$.
In Fig.~\ref{fig.3}, $Z_T Z_S^{-1} = -0.039(7) $, for the point-split current and 
$Z_T Z_S^{-1} = 0.185(7)$, for the local current \cite{feo2}. 
Also, the combination of $Z_T Z_S^{-1}$ shows no dependence on $k$. In fact, the latter is 
an $O(a)$ effect.
An estimate of $Z_T Z_S^{-1}$ at the 1-loop order, using the point split current and $\beta = 2.3$ 
is $Z_T Z_S^{-1} \equiv Z_T |_{1-loop} = -0.074$ \cite{taniguchi}.
Preliminary studies for $Z_T Z_S^{-1}$, using the local current definition, are reported in 
\cite{feo4,feo5}.
Turning to Fig.~\ref{fig.4}, the expectation is that $m_R Z_S^{-1}$ vanishes linearly when
$k \to k_c$. 
Also, a determination of $k_c$ by performing an extrapolation to zero gluino mass 
is obtained. The results is $k_c = 0.19750(38)$ 
for the point split current and $k_c = 0.19647(27)$ for the local one.
These values can be compared with the previous determination 
from the first order phase transition, $k_c = (0.1955 \pm 0.0005)$ \cite{kirchner},
see also Fig.~\ref{fig.1}.
A study of the continuum limit of the lattice SUSY WIs, using Wilson fermions, at the 1-loop order is 
\cite{taniguchi,feo5}. It is interesting to see whether an analytic calculation of the $Z_S$ and $Z_T$ 
renormalization factors is in agreement with the numerical ones.
Compared with numerical simulations \cite{feo2}, a lattice perturbative calculation 
of the SUSY WIs is more complicated. In order to do perturbation theory, we have to fix the gauge, 
which implies that new terms appear in the SUSY WIs: the gauge fixing term, the Faddeev-Popov term
and a term coming from the gauge variation of the involved operators \cite{feo5}. 
This is also discussed in \cite{dewit}. Taking into account all these contributions we have \cite{feo5},
\beeqa
&& \hspace{-0.7 cm} \big< {\cal O} \nabla_\mu S_\mu(x) \big>  -
2 m_0 \big< {\cal O} \chi(x)  \big> + 
\big< \frac{\delta{\cal O}}{\delta \bar \varepsilon(x)}|_{\varepsilon = 0} \big> - \nonumber \\ 
&& \hspace{-0.7 cm} \big< \frac{{\cal O} \delta S_{GF}}{\delta \bar \varepsilon(x)}|_{\varepsilon = 0} \big> - 
\big< \frac{{\cal O} \delta S_{FP}}{\delta \bar \varepsilon(x)}|_{\varepsilon = 0} \big> = 
\big< {\cal O} X_S(x) \big> \, . \nonumber 
\eneqa
In \cite{taniguchi,feo5} the operator insertion studied (at the 1-loop order) 
is ${\cal O} := A_\alpha^a(y)\, \bar \lambda^b(z)$, 
which is a non-gauge invariant operator (a similar choice of operators as in the numerical case, 
for example $\chi$, would lead to a 2-loop calculation). 
In this case, $X_S(x)$ mixes with operators of equal or lower dimension \cite{vladikas,bochicchio}, as for the 
on-shell case, plus $\sum_i Z_i A_i$,
where the additional operators $A_i$ correspond to mixing with non-gauge invariant operators. 
In \cite{taniguchi}, an on-shell 1-loop perturbative calculation (using an exact expression for 
$X_S$), gives
$Z_T Z_S^{-1} \equiv Z_T |_{1-loop} = -0.074$, using a point-split definition for the supercurrent,
while preliminary studies using the local supercurrent, in the off-shell regime \cite{feo5}, 
give also results of the same order for $Z_T$ \cite{alex}.

\subsection{Ginsparg-Wilson (GW) fermions}
The GW relation \cite{ginsparg} is defined to be
\beeq
\gamma_5 D + D \gamma_5  = a D \gamma_5 D \, ,
\eneq
where $\gamma_5 D$ is an hermitian lattice Dirac operator. 
An explicit solution to the GW algebra, free of doubling species is reported in \cite{neuberger2}.
It exhibits highlight chiralities properties \cite{luscher2} and locality \cite{hernandez}.
Either DWF (already discussed) or operators obtained from a renormalization group (RG) blocking 
of the continuum one, the fixed-point (FP) operator \cite{niedermayer}, satisfy the GW relation.

Theoretical applications of GW algebra to lattice supersymmetry are reported in 
\cite{fujikawa2,fujikawa,aoyama,so} for the Wess-Zumino model \cite{zumino}.
In \cite{fujikawa2}, it is suggested that a lattice version of a perturbatively finite theory 
preserves supersymmetry to all orders in perturbation theory, in the sense that the SUSY 
breaking terms induced by the failure of the Leibniz rule become irrelevant in the 
continuum limit. Differences between implementing GW fermions and Wilson fermions are 
also analyzed.
In \cite{fujikawa} a conflict between lattice chiral symmetry and the 
Majorana condition for a Yukawa-type coupling is pointed out. If one adopts GW operators, 
a precise analysis of supersymmetry and its breaking require 
a consistent formulation of Majorana fermions. 

In Ref.~\cite{itoh} a new formulation of SYM on the lattice with an exact fermionic symmetry 
is presented. 
First, it is considering the model in a fundamental lattice which is called 
the one-cell model and it is derived the preSUSY transformations. Then, it is extended
to the entire lattice.
The lattice action has a peculiar form: a $2a$ translational invariance, not the usual
$a$ ones, and it is called Ichimatsu lattice (similar to a chessboard).
At this conference, first non-perturbative results on an Ichimatsu lattice gauge theory
is presented \cite{sawanaka}, while in \cite{sawanaka2} the study of the phase transition 
is reported.
Although this is an interesting formulation, one has to note that, because of the staggered 
fermion action, there is no exact balance between fermionic and bosonic degrees of freedom.

\section{EXACT SUPERSYMMETRY ON THE LATTICE}
Improving lattice supersymmetry seems to be a difficult task for gauge theories.
In fact, most SUSY theories, as for example $N=2$ or $N=4$ SYM, contain scalar bosons which typically 
produce SUSY violating relevant operators, which has to be fined tune away. For 
$N=1$, as we have seen, only a fine tuning is needed in order to eliminate the mass term.
Unfortunately, because there is no discrete version of supersymmetry which can be implemented 
to forbid scalar masses and unwanted relevant operators,
it is desiderable to construct lattice structures which directly display at least
a subset of exact supersymmetry in order to decrease the number of fine tuning to do.
In Ref.~\cite{catterall} a similar approach is used, for the $2d$ 
Wess-Zumino model with extended $N=2$ supersymmetry on the lattice. The model is then discretized 
in a way that preserves exactly a subset of the continuum SUSY transformations.
This is enough to guarantee that the full symmetry is restored without fine tuning in the 
continuum limit. Also numerical results using the HMC algorithm are presented \cite{catterall}, and show 
the equality between fermions and boson masses and also the verification of the WIs to high precition.

Another nice example of exact lattice supersymmetry is \cite{bietenholz},
where a perfect SUSY action for the $2d$ and $4d$ Wess-Zumino model, free case,
is presented. The perfect action is achieved in terms of block variable 
RG transformation, which maps a system from a fine lattice to a 
coarser lattice in a specific way, that keeps the partition function and all expectation values invariant.
It is preserved invariance under a continuous SUSY type of field transformations in a local
perfect lattice action, which contains also a remnant chiral symmetry. This perfect formulation
also cures the problem related with the Leibniz rule on the lattice \cite{dondi}.
In the perfect lattice formulation it is obtained the consistently blocked continuum translation
operator. Therefore the algebra with the field variations closes. 
Ref.~\cite{so} uses a similar approach.

At this conference a talk by Kaplan \cite{kaplan3}, 
based on a recent remarkable paper \cite{kaplan4} has been presented. 
It is a new method for implementing supersymmetry on a spatial 
lattice for a variety of SYM theories in $3+1$ dimensions, including $N=4$, 
in a way that eliminates or reduces the problem of fine tuning. 
The formalism is presented in the Minskowski space but a generalization to 
Euclidean space is under way.
The motivation is based on a recent work on deconstruction of SUSY theories \cite{arkani}.
The spatial lattice is created by ``orbifolding''.
Starting from a ``mother theory'', being a SUSY quantum mechanics (QM) system with extended supersymmetry,
a gauge group $U(k N^d)$, and a global $R$ symmetry group $G_R$, the spatial lattice is 
constructed by orbifolding out a $Z_N^d$ factor from $G_R \times U(k N^d)$, which will result in a 
$U(k)$ gauge symmetry, living on a $d$-dimensional spatial lattice. 
This is called the ``daughter theory''.
Taking then the continuum limit of this ``daughter theory'', in some point in the classical moduli space of 
vacua, produces a higher dimensional quantum field theory with the original extended supersymmetry restored 
(previoulsy broken by the orbifolding procedure), together with Poincar\'e invariance.
The important point is that the lattice generated by orbifolding can retain some 
of the exact supersymmetries which facilitates the recovery of the remaining ones 
in the continuum limit.
This method also protects the renormalizability of the theory.
In \cite{kaplan4}, a SUSY QM model with extended supersymmetry in $0+1$ dimensions is considered, and then 
the types of lattices that can be obtained via orbifolding are shown.

\subsection{Other related topics}
At this conference, a nice example of a Hamiltonian lattice version for the $2d$ Wess-Zumino
model is presented by Campostrini \cite{feo6}. Previous results in \cite{feo7}.
Developing numerical simulations techniques using the Hamiltonian approach \cite{kogut}
turns out to be very advantageous (but contains also some disadvantages). 
Powerful many-body techniques are available: the Green Function Monte Carlo (GFMC) algorithm \cite{linden}.
Since $H$ is conserved, it is possible to preserve exactly a $1d$ SUSY subalgebra of 
the continuum $N=1$ SUSY algebra. This is an advantage in comparison with the standard lattice 
formulation and is it enough to guarantee the most important properties of SUSY-like paring of 
positive energy states \cite{feo6,feo7}. 
Moreover, fermions are implemented directly, and need not be integrated out.
A disadvantage is that fermions may lead to sign problems, but
at least in $1+1$ dimensions, it can be by-passed.
In \cite{feo6}, simulations using GFMC are performed. The algorithm is very efficient in computing the ground
state energy $E_0$ (because it can distinguish between 0 and $10^{-5}$), and therefore can be used to study 
the pattern of supersymmetry breaking, which corresponds to $E_0 > 0 $, while unbroken supersymmetry 
corresponds to $E_0 =0$.
Focusing on $V(\phi) = \lambda \phi^2 + \lambda_0$, predictions for the strong coupling and the 
weak coupling regime are
quite different and it is interesting to study both numerically and analytically and the crossover from
the strong to the weak coupling \cite{feo6}.

At this conference a related and very interesting method, namely, SUSY Discretized Light-Cone Quantization (SDLCQ)
has been reported by Trittmann \cite{trittmann}. SDLCQ is a discrete, Hamiltonian and manifestly SUSY 
approach to solving a quantum field theory. Some results (including a review) 
for the $N=1$ SUSY Chern Simons theory, 
in $2d$ and $3d$, including correlators and bound states are in \cite{pinsky}. 
The theory is discretized by imposing periodic boundary conditions on the boson
and fermion field in terms of discrete momentum modes, $k = n P/ K $, where 
$K$ is a positive integer that determines the resolution (typically of order 10 in the numerical calculations). 
The continuum limit is reached for $K \to \infty$. 
The low energy spectrum mass is determined using the fit $M^2 = M_\infty^2 + b(1/K)$ and 
approximate BPS states which have non-zero masses are found.
In \cite{trittmann}, a numerical test of the Maldacena conjecture within the $10-15 \%$ is found.
A schematic comparison of SDLCQ and lattice gauge theory is also presented. 

Ref.~\cite{wosiek} casts a Hamiltonian study of SYM QM,
which faces the problem of the introduction of a cut-off that violates supersymmetry 
and the restoration of supersymmetry when it is taken away.
It has been applyed to the Wess-Zumino QM and for the SYM QM, in $2d$ and $4d$. There are no sign problems.
Also, the complete spectrum, Witten index and identification of SUSY multiplets has been 
determined \cite{wosiek2}. 

\section{CONCLUSIONS}
A big effort has been made in order to describe supersymmetry on the lattice.
Traditional Wilson fermions have been used in realistic computations with nice results. 
Improved chiral fermions results are starting too. New exciting ideas \cite{kaplan4} are now 
waiting for numerical applications.
It has also been shown that most properties of $N=1$ SYM are analogous to 
non-SUSY YM and QCD and can be tested on the lattice 
in order to understand a possible transition from SYM to YM.
For this reason, placing real-world QCD in a wider variety of theories may help us to better understand it.
\vskip2mm
\noindent
{\bf Acknowledgments.} It is a pleasure to thank M.~Beccaria, M.~Campostrini, F.~Farchioni, T.~Galla, C.~Gebert, 
R.~Kirchner, M.~L\"uscher, I.~Montvay, G.~M\"unster, J.~Negele, R.~Peetz and A.~Vladikas.
Also, W.~Bietenholz, M.~Golterman, D.~B.~Kaplan, C.~Pena, S.~Pinsky, H.~So, U.~Trittmann and
J.~Wosiek, for discussions and private communication. 
M.~Peardon and S.~Ryan for the nice and stimulating atmosphere at Trinity College.
This work was partially funded by the Enterprise-Ireland grant SC/2001/307.

\end{document}